\documentclass{webofc}
\usepackage[varg]{txfonts}   

\wocname{EPJ Web of Conferences}

\woctitle{INPC 2013}

\begin{document}
\title{Nuclear lattice simulations: Status and perspectives\footnote{Lecture sponsored by
    The European Physical Journal {\bf A} -- Hadrons and Nuclei}}

\author{Ulf-G. Mei{\ss}ner\inst{1,2}\fnsep\thanks{\email{meissner@hiskp.uni-bonn.de}} 
}

\institute{Universit\"at  Bonn, HISKP \& BCTP, D-53115 Bonn, Germany
\and
           Forschungszentrum J\"ulich, IAS, IKP \& JCHP, D-52425 J\"ulich, Germany 
          }

\abstract{%
  I review the present status of nuclear lattice simulations.\\
  This talk is dedicated to the memory of Gerald E. Brown.
}
\maketitle
\section{Prologue: In memoriam of Gerald E. Brown}
\label{pro}
Just a few days before this talk, my beloved teacher Gerry Brown passed away
in the age of 86. Not only me but also the nuclear physics community owes him a
lot. We will always remember him as a superb physicist, a gifted teacher and
a wonderful (hu)man. Thank you, Gerry!

\section{Introduction}
\label{intro}

This talk is a natural extension of my presentation at INPC 2004, where I gave a
plenary talk titled ``Modern theory of nuclear forces''
\cite{Meissner:2004yy}. There, I presented the developments of the chiral
effective field theory (EFT) approach to the nuclear force problem initiated
by Steven Weinberg in 1990~\cite{Weinberg:1990rz}. This scheme allows for
a consistent and accurate description of the forces between two, three and
four nucleons, based on the symmetries of QCD, with controllable theoretical
errors and a systematic scheme to improve the precision. Much progress has
been made in the applications and tests of these
forces in few-nucleon systems, for a cornucopia of  bound state or scattering
calculations, see e.g. the recent  reviews
\cite{Epelbaum:2008ga,Machleidt:2011zz} 
and also  the talk by Machleidt at this conference. It is therefore natural to
ask how to address the properties of nuclei with $A>4$. There are essentially
two ways of doing that. On the one hand, one can combine
standard many-body techniques like the (no-core) shell model or coupled
cluster approaches with these forces as discussed e.g by Bacca and Roth (often
using a low-momentum a.k.a. soft representation of the forces). On the
other hand, one can try to devise a new approach to tackle the many-body
problem that is tailored to the chiral EFT approach. This novel scheme, that combines
the EFT description of the few-nucleon forces with Monte-Carlo simulation
techniques, will be discussed here. It is called {\bf nuclear lattice simulations} 
and was introduced for atomic nuclei in  Ref.~\cite{Borasoy:2006qn} and
subsequently developed in few-nucleon systems (see the review by Lee that
contains references to earlier related and pioneering work \cite{Lee:2008fa}).
Here, I will discuss the progress made in the description of nuclei that has
been  made in the last years, particularly with respect to the spectrum and
structure of $^{12}$C and other $\alpha$-cluster type nuclei.

\section{Chiral effective field theory on a lattice}
\label{sec-1}

The starting point of chiral EFT for nuclei is the standard effective pion-nucleon
Lagrangian supplemented by multi-nucleon field operators, that faithfully
reproduces the chiral Ward identities of QCD. The latter terms are not
constrained by chiral symmetry and their parameters (low-energy constants,
called LECs for brevity) must be determined from a fit to  few-nucleon
data. Following Weinberg, the power counting is performed on the level of the
effective potential between two, three and four nucleons and a properly
regularized Schr\"odinger-type equation is solved to generate the bound and
scattering states. Life becomes much simpler if one goes to Euclidean time and
formulates the theory on a space-time lattice of volume $(N\, a)^3 \times
(N_t \, a_t)$, with $a\, (a_t)$ the (temporal) lattice spacing and $N\, (N_t)$ the 
number of grid points in the space (time) direction. In fact, the lattice
spacing $a$ serves as an UV regulator of the EFT as it implies a maximal
momentum  of $p_{\rm max} \sim \pi/a$. For a typical value of $a \simeq
2\,$fm, one has  $p_{\rm max} \simeq 300\,$MeV, i.e. a very soft interaction.
A further advantage of the lattice formulation is that is automatically takes
into account {\bf all} possible configurations allowed by symmetries, in
particular, we can have up to four nucleons on one lattice site. Such
configurations naturally lead to clustering, in fact, they require some
special treatment so as not to produce overbinding of $\alpha$-like
configurations. For that reason, the leading order (LO) interaction is smeared
with a Gaussian, where the width parameter is adjusted to give the proper
average S-wave NN effective range. Such  smeared sources are also
used frequently in lattice QCD calculations of hadron properties and dynamics.
This also means that the LO lattice EFT result should not be directly
compared with a corresponding continuum calculation, where such effects are
generated by higher order operators in the full potential. In the lattice
approach, only the LO potential is summed up whereas all higher orders are
treated in perturbation theory. This is consistent with the low momentum
cutoff utilized. An important new development is the so-called {\em nuclear lattice
 projection Monte Carlo (MC) technique} \cite{Epelbaum:2012qn}. 
We use a larger class of initial and final states than considered in our
earlier  work. I give some examples: For the calculation of
$^{4}$He we use an initial state with four nucleons, each at zero momentum. 
For the calculation of $^{8}$Be we use the same initial state as for $^{4}$He,
but then apply creation operators after the first time step to inject four
more nucleons at zero momentum.  The analogous process is done to extract
four nucleons before the last step. 
\begin{figure}[htb!]
\centering
\sidecaption
\includegraphics[width=6.75cm,clip]{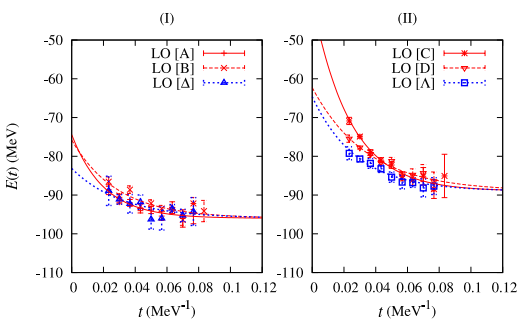}
\caption{Lattice results for the $^{12}$C spectrum at leading order. In Panel
I results from three different initial states, A, B, and $\Delta$ are shown,
each approaching the ground state energy. In Panel II  results starting
from three other initial states, C, D, and $\Lambda$ are displayed. These trace out an
intermediate plateau at energy about $7$~MeV above the ground state. This is
nothing but the celebrated Hoyle state.}%
\label{fig-gs}       
\end{figure}
This injection and extraction process of
nucleons at zero momentum helps to eliminate directional biases caused by
initial and final state momenta. To probe the structure of  the $^{12}$C
states, we use not only delocalized initial and final states (configurations
A,B,C,D) but also cluster-type wave functions (compact triangle $\Delta$,
elongated triangle $\Lambda$). As seen in Fig.~\ref{fig-gs}, while the
configurations A,B and $\Delta$ approach the ground state energy, we do not
find fast convergence to the ground state for configurations C,D and
$\Lambda$, thus giving the structure of the first excited $0^+$ state.  
The common thread connecting each of the initial states C, D, and $\Lambda$, 
is that each produces a state which has an extended or prolate geometry.
The projection MC method is thus a powerful tool to unravel the structure of
any considered nucleus and its excited states.

\section{Fixing parameters and ground state results}
\label{sec-2}

Next, we must fix the parameters (LECs) of our theory. At present, we
work at next-to-next-to-leading order (NNLO) in the chiral expansion, including
strong and electromagnetic isospin breaking (the Coulomb force is included
as described in \cite{Epelbaum:2010xt}). We have 9 strong 
isospin-symmetric LECs. We fix  these from the S- and P-waves in $np$ scattering
and the deuteron quadrupole moment, $Q_d$. The 2 LECs related to
isospin-breaking are determined from a fit to the different $np$, $pp$ and
$nn$ scattering lengths, whereas the two LECs $C$ and $D$ of the leading
three-nucleon force are fixed from the $nd$ doublet scattering length 
and the triton binding energy. The first non-trivial predictions are
then the momentum dependence of the $pp$ $^1S_0$ partial wave, which agrees
nicely with the Nijmegen PWA result \cite{Epelbaum:2010xt} and the binding 
energy difference of the triton and $^3$He. We find $E(^3{\rm He}) - E(^3{\rm
  H}) = 0.78(5)\,$MeV \cite{Epelbaum:2009pd}, in good agreement with the
empirical value of $0.76\,$MeV.

The ground state energies for various nuclei are summarized in
Tab.~\ref{tab-gs}. Consider first $^{4}$He, $^{8}$Be and $^{12}$C,
which show some unique trends. We observe that the smeared LO contribution
is already close to the empirical numbers for these nuclei, whereas the 
two-nucleon forces alone underbind at NNLO. The rescue comes from the three-nucleon forces
at this order, which provide just the right amount of binding. These 3N forces
become more important as the atomic number $A$ increases, as is also
observed in e.g.  Greens function MC \cite{Pieper:2007ax} or
no-core shell model calculations \cite{Roth:2011vt}. The NNLO results for 
these nuclei are already quite satisfactory. Of course, higher orders 
and improved methods to extract the energies will have to be considered in the
future. Work along these lines is in progress.
The situation is somewhat different for $^{16}$O (note that these are
preliminary numbers which might still change a bit). Here, the smeared
leading order leads to some overbinding, which is not completely corrected
from the 3N forces at NNLO. I will come back to this issue in
Sec.~\ref{sec-heavy}.
\begin{table}[htb!]
\centering
\caption{Ground state energies. LO denotes the smeared leading order,
NNLO (XN) the contribution from X=2,3 nucleon forces up-to-and-including
NNLO and NNLO the total result at that order. Also, Exp. denotes the empirical
values. The numbers for $^{16}$O are preliminary. All energies in MeV.} 
\label{tab-gs}       
\begin{tabular}{|l|cccc|l|}
\hline
nucleus   &  LO   & NNLO (2N )& NNLO (3N) & NNLO & Exp. \\
\hline
$^{4}$He  & $-28.0(3)$ & $-24.9$ & $-3.4$ &  $-28.3(6)$ & $-28.3$ \\
\hline
$^{8}$Be  & $-55(2)$ & $-48$ & $-7$ &  $-55(2)$ & $-56.5$ \\
\hline
$^{12}$C  & $-96(2)$ & $-77$ & $-15$ &  $-92(3)$ & $-92.2$ \\
\hline
$^{16}$O  & $-147.5(5)$ & $-121$ & $-20$ &  $-141.4(9)$ & $-127.6$ \\
\hline
\end{tabular}
\end{table}

\section{Spectrum and structure of carbon-12}
\label{sec-C12}

The spectrum of $^{12}$C is one of the most challenging topics in nuclear
theory. More precisely, the so-called Hoyle state, a $0^+$ excitation about
7.7~MeV above the ground state, has been an enigma to {\em ab initio}
calculations since it was predicted by Hoyle in 1954 \cite{Hoyle:1954zz}
from a calculation of the 
abundance of carbon and oxygen through stellar fusion and experimentally
verified at Caltech a few years later \cite{Cook:1957zz}. Before continuing,
I have to define what is precisely meant by an {\em ab initio} calculation.
In a first step, one fixes the underlying forces between two and three
nucleons in few-body systems, i.e. bound and scattering observables with 
$A \leq 4$. Then, given the so determined forces, one solves the quantum
$A$-body problem exactly, in this case using lattice Monte Carlo 
methods\footnote{The interesting and successful fermion molecular
dynamics calculation of Ref.~\cite{Chernykh:2007zz} requires a fit to a broad
range of nuclei to pin down the various model parameters, see also the
contribution by Feldmeier and Neff to this conference~\cite{Feldmeier:2013qta}.}.
Any nuclear property for systems with $A\geq 5$ is then a prediction, the
precision of the calculations can be systematically improved by going to
higher order in the potential and improving the statistical and systematic
errors of the MC signals. As with any new method, it is only accepted by the
community if one is able to solve problems that could not be resolved before.
That is why as a first application of the framework of nuclear lattice
simulations we focused on the spectrum of  $^{12}$C. In fact, based on 
standing waves properly projected onto zero total momentum as well as 
angular momentum and parity, the
first ever ab initio calculation of the Hoyle state was reported 
in~\cite{Epelbaum:2011md}. Since then, using the projection MC method,
we have improved the calculations and gained further insight into the
structure of the ground and excited state, as summarized in Tab.~\ref{tab-C12}.
\begin{table}[htb!]
\centering
\caption{Low-lying spectrum of $^{12}$C at NNLO from 2N and 3N forces
\cite{Epelbaum:2012qn} in comparison to experiment.}
\label{tab-C12}       
\begin{tabular}{|c|cccc|}
\hline
        & $0_1^+$   & $2_1^+$ & $0_2^+$ & $2_2^+$  \\
\hline
2N  & $-77$ & $-74$ & $-72$ &  $-70$  \\
3N  & $-15$ & $-15$ & $-13$ &  $-13$  \\
2N+3N  & $-92(3)$ & $-89(3)$ & $-85(3)$ &  $-83(3)$  \\
\hline
Exp.  & $-92.16$ & $-87.72$ & $-84.51$ &  $-82.6(1)$ \cite{Freer:2009zz} \\
      &          &          &          &  $-82.13(11)$ \cite{Zimmerman:2013cxa} \\
\hline
\end{tabular}
\end{table}
The trends for the excited states are similar to the ground state - about 20\%
of binding from the 3N forces is required to bring agreement between
experiment and theory. Of particular interest besides the Hoyle state is 
the $2^+$ excitation just 2~MeV above it. This state can be interpreted as an
excitation of the elongated triangle that features prominently in the
structure of the Hoyle state. The predicted energy is in good agreement
with recent measurements, see Tab.~\ref{tab-C12} (cf. also the
group-theoretical anlysis of such cluster states in Ref.~\cite{Bijker:2000fw}).
In
Ref.~\cite{Epelbaum:2012qn}, electromagnetic observables (charge radii,
quadrupole moments and electromagnetic transitions among the even-parity
states of $^{12}$C) were also given at LO - all these results are in
reasonable agreement with experiment. However, the calculation of the higher
order corrections turned out to be more complicated than the ones for the
spectrum. This requires algorithmic improvements for a precise extraction
of the corresponding matrix elements which are presently under
investigation.

\section{Life on Earth - an accident?}
\label{sec-life}

\begin{figure}[htb!]
\centering
\sidecaption
\includegraphics[width=6.75cm,clip]{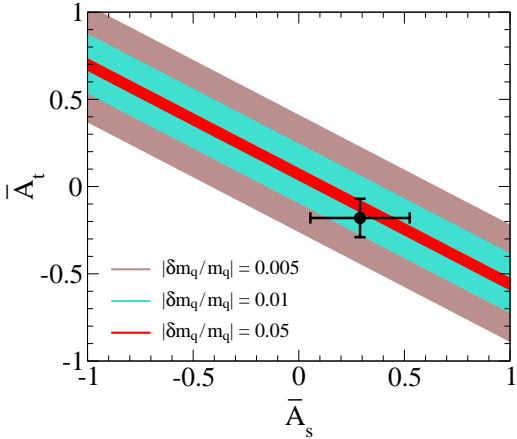}
\caption{End-of-the-world plot. For relative quark mass changes
of 0.5\% (wide band), 1\% (middle band) and 5\% (inner most band), one
can read off the allowed values for  $\bar A_{s}^{}$ and $\bar A_{t}^{}$ (both
varied within natural ranges from $-1$ to $+1$),
that are consistent with the requirements \cite{Schlattl:2003dy} 
for car\-bon-oxygen based life. 
The recent determination of $\bar A_{s}^{}$ and $\bar A_{t}^{}$ from
Ref.~\cite{Berengut:2013nh} is depicted by the black cross.
}
\label{fig-end}       
\end{figure}

The Hoyle state has been heralded as a prime example of the anthropic
principle\footnote{See, however, Kragh~\cite{Kragh} for a critical view of his issue.}, 
which states that the parameters of the fundamental interactions
take values that are compatible with the existence of carbon-based life and
that the Universe is old enough for such life to have formed (for a recent
review, see \cite{Schellekens:2013bpa}). {\em Computer simulations} allow one to 
address the question of how sensitive the production of carbon and oxygen in 
hot stars is to the fundamental parameters of QCD+QED,
the gauge field theories underlying the formation of atomic nuclei.
This is one of the great strengths of the framework presented here, as
I will outline in the following. In fact, the energy of the Hoyle state 
is just 379~keV away from the $3\alpha$ threshold,
and the production rate of carbon in the $3\alpha$ process scales with
$\exp(-\Delta E/k_BT)$, with $\Delta E = E_{\rm Hoyle}-3E_\alpha$, $T$ is the
stellar temperature and $k_B$ Boltzmann's constant.  In earlier studies,
it was shown that $\Delta E$ can be modified by $\pm 100\,$keV, so that
one still has a sufficient production of carbon and oxygen
\cite{Schlattl:2003dy}. This does not appear particularly fine-tuned, as
variation of about 25\% are consistent with carbon-oxygen based life. However,
we have to translate this constraint on $\Delta E$ into constraints on the
fundamental parameters of QCD+QCD, here the average light quark mass $m_q$ and
the fine-structure constant $\alpha_{\rm QED}$, see the recent works for more
details \cite{Epelbaum:2012iu,Epelbaum:2013wla}. Note also that the 
strong coupling constant $\alpha_S$
is fixed from scale-setting through the nucleon mass, so it can not be varied 
independently. In the following, we only consider small changes in $m_q$ and
$\alpha_{\rm QED}$, so that 
$\delta E_i^{} \simeq ({\partial E_i^{}}/{\partial M_\pi^{}})|_{M_\pi^{\rm
 ph}} \delta M_\pi^{} + \left.({\partial E_i^{}}/{\partial \alpha_\mathrm{QED}^{}}) 
\right. |_{\alpha_\mathrm{QED}^\mathrm{ph}} \delta \alpha_\mathrm{QED}^{}$, where
the superscript 'ph' refers to the physical value and we use the 
Gell-Mann--Oakes--Renner relation $M_\pi^2 \sim m_q$ to translate the
quark mass dependence into the pion mass dependence. The knowledge of the
pion mass dependence of the nuclear Hamiltonian is required to calculate
the pertinent derivatives. In fact, consider any nuclear energy level or
difference under small perturbations of the pion mass around its empirical
value to leading order in the chiral expansion,
\begin{equation}\label{delE}
\left. \frac{\partial E_i^{}}{\partial M_\pi^{}} \right|_{M_\pi^\mathrm{ph}} =
\left. \frac{\partial E_i^{}}{\partial\tilde M_\pi^{}} \right|_{M_\pi^\mathrm{ph}}
+ x_1^{} \left. \frac{\partial E_i^{}}{\partial m_N^{}} \right|_{m_N^\mathrm{ph}}
+ x_2^{} \left. \frac{\partial E_i^{}}{\partial g_{\pi N}^{}} 
\right|_{g_{\pi N}^\mathrm{ph}} 
+ x_3^{} \left. \frac{\partial E_i^{}}{\partial C_0^{}}
\right|_{C_0^\mathrm{ph}}
+ x_4^{} \left. \frac{\partial E_i^{}}{\partial C_I^{}} \right|_{C_I^\mathrm{ph}},
\end{equation}
where $\tilde M_\pi^{}$ refers to the explicit $M_\pi^{}$-dependence from the 
pion propagator in the OPE contribution, and
$x_1 = (\partial m_N/\partial M_\pi)|_{M_\pi^{\rm ph}}$  and 
$x_2  = (\partial g_{\pi N})/\partial M_\pi)|_{M_\pi^{\rm ph}} = 
(1/2F_\pi)  (\partial g_A/\partial M_\pi)|_{M_\pi^{\rm ph}} 
- (g_A/2F_\pi^2)(\partial F_\pi/\partial M_\pi)|_{M_\pi^{\rm ph}}$ 
are obtained from the quark mass dependence of the
nucleon mass and the nucleon axial-vector coupling combined with the one of
the pion decay constant utilizing lattice QCD data and chiral perturbation theory.
The coefficients $x_3,x_4$ can be mapped uniquely on the quark mass dependence
of the inverse nucleon-nucleon S-wave scattering lengths, denoted 
$\bar A_{s}^{} \equiv {\partial a_{s}^{-1}}/{\partial M_\pi^{}}
|_{M_\pi^{\rm ph}}$ and
$\bar A_{t}^{} \equiv {\partial a_{t}^{-1}}/{\partial M_\pi^{}}
|_{M_\pi^{\rm ph}}$. These can be determined from chiral nuclear EFT,
assuming some modeling based on resonance saturation \cite{Epelbaum:2001fm}.
The most recent analysis at NNLO gives $\bar A_{s}^{} = 0.29_{-0.23}^{+0.25}$,
$\bar A_{t}^{} = -0.18 \pm 0.10$ \cite{Berengut:2013nh}. This introduces 
some amount of uncertainty, as discussed below. The various partial
derivatives in Eq.~(\ref{delE}) can be straightforwardly and precisely computed
using the auxiliary field quantum Monte Carlo techniques (for details, see
Ref.~\cite{Epelbaum:2013wla}). Independent of the precise values of 
 $\bar A_{s}^{}$ and $\bar A_{t}^{}$, we find strong evidence that 
the physics of the $3\alpha$-process is driven by $\alpha$-clustering,
as speculated before (see e.g. \cite{Weinberg}). The condition that 
$|\Delta E| \leq 100\,$keV can be turned into an inequality that connects 
the quark mass change $\delta m_q/m_q$ with $\bar A_{s}^{}$ and 
$\bar A_{t}^{}$, i.e. $(c_1\bar A_{s}^{} + c_2\bar A_{t}^{} + c_3) \cdot (\delta m_q/m_q) \leq
|\Delta E|$ (where the $c_i$ are calculated using our MC techniques), 
as depicted in Fig.~\ref{fig-end}, the so-called 
{\em end-of-the-world plot}. From that figure one sees that shifts 
in the light quark mass at the $\simeq 2 - 3 \%$ level are unlikely 
to be detrimental to the development  of life. Tolerance against 
much larger changes cannot be ruled out at present, given the relatively 
limited knowledge of the quark mass dependence of the two-nucleon S-wave 
scattering parameters $\bar{A}_{s,t}$. Lattice QCD is
expected to provide refined estimates of the scattering parameters in the future.
Further, variations of the fine-structure constant $\alpha_{\rm QED}$ up to
$\pm 2.5\%$ are consistent with the requirement of sufficient carbon and
oxygen production in stars.

\section{Towards heavier nuclei}
\label{sec-heavy}

\begin{figure}[htb!]
\centering
\sidecaption
\includegraphics[width=7.75cm,clip]{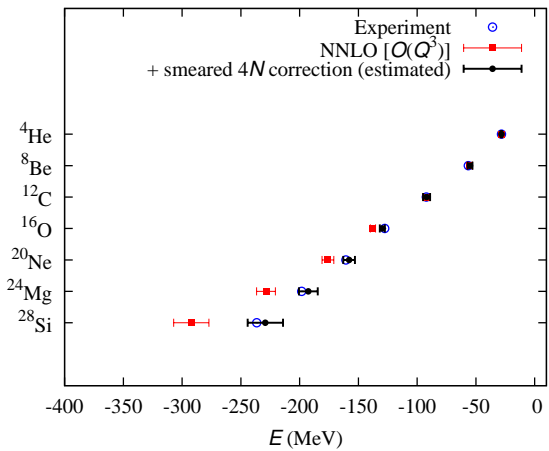}
\caption{Ground state energies for $\alpha$-cluster nuclei at NNLO
(red boxes) in comparison to the empirical numbers (blue open circles).
Including a smeared four-nucleon contact interaction with its strength fixed
from the $^{16}$O ground state energy leads to much improved results (black circles).
The numbers (in MeV) are $158(5) \,[-160.6]$,  $193(8) \,[-198.3]$ and  $229(15)
\,[-236.5]$ for  $^{20}$Ne, $^{24}$Mg, and $^{28}$Si, respectively, and the
value of the experimental binding energy is given in the square brackets.
}
\label{fig-alpha}       
\end{figure}

Since the formalism outlined above does particularly well for $\alpha$-cluster 
states (note, however, that it is not restricted to such type of nuclei), 
it is natural to ask what happens if one extends these simulations to the next few
$\alpha$-cluster nuclei, that are $^{16}$O,  $^{20}$Ne, $^{24}$Mg, and
$^{28}$Si. Here, I report some preliminary numbers for the ground state energies
and the low-lying spectra of these nuclei. First, it is interesting to note 
that for this mass range
the CPU time scales approximately as $A^{2}$ (whereas naively one would 
expect an $A^3$-scaling from the determinant calculation), so that the 
required computing time for 
$^{28}$Si is only 5.4 times the one for the $^{12}$C ground state. Second,
we find that these heavier nuclei are overbound at NNLO, with the overbinding
increasing with mass number, cf. Fig.~\ref{fig-alpha}. Let us consider the 
 $^{16}$O nucleus. Using the same action as described before leads to 
$E_{\rm g.s.} (^{16}{\rm O}) = (-141\pm 1)\,$MeV, which is about 14~MeV
overbinding (for an earlier variational MC calculation based on non-chiral
forces, see~\cite{Pieper:1992gr}). This effect 
can be traced back to the new $\alpha$-cluster geometries (the square and
the tetrahedral structure, to be precise) that are possible
with four or more alpha particles. These geometries generate too much
attraction, similar to the four-nucleon configurations on one lattice site 
that required Gaussian smearing as discussed above. The simplest method to 
overcome  this effect is to introduce a smeared four-nucleon operator, which is 
formally of higher order in the chiral expansion (for an earlier use of such 
type of operator, see e.g. Refs.~\cite{Epelbaum:2009pd,Epelbaum:2010xt}).
The strength of this operator can be fixed from the ground state energy of
$^{16}$O. As can be seen from Fig.~\ref{fig-alpha}, this also leads to a good
description of the ground state energies of {\em all} $\alpha$-cluster nuclei
up to $^{28}$Si. Therefore, we believe that we have captured correctly the
physics behind the binding energies of these nuclei. Clearly, these studies need
to be backed by higher order calculations. In fact,  we are presently extending the 
chiral potentials to next-to-next-to-next-to-leading order (N$^3$LO). These exist 
in the continuum (see e.g. the N$^3$LO three-nucleon forces in 
Refs.~\cite{Ishikawa:2007zz,Bernard:2007sp,Bernard:2011zr}) and just need to
be adopted to the lattice notation.

It is also interesting to study the 
low-lying spectrum of these nuclei. Consider again $^{16}$O.
The tetrahedral structure is the dominant component in its ground state wave
function with $J^\pi = 0^+$ and by its geometry it naturally leads to a $3^-$ excitation. The
square configuration is dominant in the first $0^+$ excitation, that is located
just below the $3^+$-state and has a $2^+$ configuration as its lowest
excitation. These configurations are exactly matching the experimentally
found low-lying states in  $^{16}$O (see also~\cite{xx}). In fact, we are presently performing a
detailed analysis of this spectrum and the influence of the breaking of
rotational symmetry of such configurations on the  coarse lattices
with $a \simeq 2\,$fm used.

\section{Summary and outlook}
\label{sec-summ}

To summarize, nuclear lattice simulations are a novel scheme to tackle the
nuclear quantum many-body problem, firmly rooted in the so successful chiral effective 
field theory of the nuclear forces. To make further progress, methodological 
investigations in three directions are required. First, one has to address the
topic of corrections beyond NNLO. Second, systematic studies of the lattice
spacing dependence and the related issue of rotational symmetry breaking are
required. Third, the hybrid MC algorithm has to be improved to get better
signals for electromagnetic and weak operators. Work in all these directions
is under way. Finally, I mention a few  physics topics that are presently 
also under investigation:
\begin{itemize}
\item {\em Reaction theory on the lattice.} A general strategy for investigating 
capture reactions on the lattice as been outlined in
Ref.~\cite{Rupak:2013aue}. The method consists of two major steps. First, 
projection MC is used to determine a multi-channel adiabatic lattice
Hamiltonian for the participating nuclei. The second part is the calculation
of the effective two-body capture reaction at finite volume using the adiabatic 
Hamiltonian. As a first application, the leading M1 contribution for
$p(n,\gamma)d$ is worked out in~\cite{Rupak:2013aue}. See also 
Ref.~\cite{Navratil:2011zs} for an ab initio approach for fusion reactions
using chiral forces.
\item{\em Topological volume corrections.} Scattering processes on a lattice 
can be analyzed using L\"uscher's method \cite{Luscher:1990ux}. 
However, if the bound states
(like nuclei) move in a periodic volume, additional topological volume
corrections arise that are  sensitive to the number and mass of the constituents  
\cite{Bour:2011ef} (and its extension to QFT in \cite{Davoudi:2011md}).
The importance of these corrections has been  explicitely demonstrated 
in Ref.~\cite{Bour:2012hn} for elastic scattering between a fermion and 
a bound dimer in the shallow  binding limit (and is presently investigated
in the study of $nd$ scattering \cite{Rokash}).
\item {\em Equation of state of neutron matter.} The equation of state of
  neutron matter is currently a hot topic, see Steiner's talk at this
  conference. We are presently performing a NNLO analysis of the energy
  of neutron matter as a function of density,  extending the
  work in Ref.~\cite{Epelbaum:2008vj}, investigating also the pairing gap
  and possible P-wave pairing. For related works employing chiral EFT, see
  e.g. Refs.~\cite{Kaiser:2001jx,Lacour:2009ej,Gezerlis:2013ipa}.

\end{itemize}

\subsection*{Acknowledgments}

I would like to thank my colleagues from the NLEFT collaboration for sharing
their insight into the topics presented here, especially Dean Lee and Timo
L\"ahde.  This work is supported in part by DFG and NSFC (SFB/TR 110
``Symmetries and the emergence of structure in QCD''), the HGF (Nuclear
Astrophysics Virtual Institute, VH-VI-417), the BMBF (grant no. 05P12PDFTE)
and by the EU (HadronPhysics3, grant no. 283286).



\begin{thebibliography}{}

\bibitem{Meissner:2004yy} 
  U.-G.~Mei{\ss}ner,
  Nucl.\ Phys.\ A {\bf 751}, 149 (2005)

\bibitem{Weinberg:1990rz} 
  S.~Weinberg,
  Phys.\ Lett.\ B {\bf 251}, 288 (1990)


\bibitem{Epelbaum:2008ga} 
  E.~Epelbaum, H.-W.~Hammer and U.-G.~Mei{\ss}ner,
  Rev.\ Mod.\ Phys.\  {\bf 81}, 1773 (2009)

\bibitem{Machleidt:2011zz} 
  R.~Machleidt and D.~R.~Entem,
  Phys.\ Rept.\  {\bf 503}, 1 (2011)

\bibitem{Borasoy:2006qn} 
  B.~Borasoy, E.~Epelbaum, H.~Krebs, D.~Lee and U.-G.~Mei{\ss}ner,
  Eur.\ Phys.\ J.\ A {\bf 31}, 105 (2007)

\bibitem{Lee:2008fa} 
  D.~Lee,
  Prog.\ Part.\ Nucl.\ Phys.\  {\bf 63}, 117 (2009)

\bibitem{Epelbaum:2012qn} 
  E.~Epelbaum, H.~Krebs, T.~A.~L\"ahde, D.~Lee and U.-G.~Mei{\ss}ner,
  Phys.\ Rev.\ Lett.\  {\bf 109}, 252501 (2012)

\bibitem{Epelbaum:2010xt} 
  E.~Epelbaum, H.~Krebs, D.~Lee and U.-G.~Mei{\ss}ner,
  Eur.\ Phys.\ J.\ A {\bf 45}, 335 (2010)

\bibitem{Epelbaum:2009pd}
  E.~Epelbaum, H.~Krebs, D.~Lee and U.-G.~Mei{\ss}ner,
  Phys.\ Rev.\ Lett.\  {\bf 104} (2010) 142501

\bibitem{Pieper:2007ax} 
  S.~C.~Pieper,
  Riv.\ Nuovo Cim.\  {\bf 31}, 709 (2008)

\bibitem{Roth:2011vt} 
  R.~Roth, S.~Binder, K.~Vobig, A.~Calci, J.~Langhammer and P.~Navratil,
  Phys.\ Rev.\ Lett.\  {\bf 109}, 052501 (2012)

\bibitem{Hoyle:1954zz} 
  F.~Hoyle,
  Astrophys.\ J.\ Suppl.\  {\bf 1}, 121 (1954)

\bibitem{Cook:1957zz}
  C.~W.~Cook, W.~A.~Fowler, C.~C.~Lauritsen and T.~Lauritsen,
  Phys.\ Rev.\  {\bf 107} (1957) 508.

\bibitem{Chernykh:2007zz} 
  M.~Chernykh, H.~Feldmeier, T.~Neff, P.~von Neumann-Cosel and A.~Richter,
  Phys.\ Rev.\ Lett.\  {\bf 98}, 032501 (2007)

\bibitem{Feldmeier:2013qta} 
  H.~Feldmeier and T.~Neff,
  arXiv:1307.6449 [nucl-th]

\bibitem{Epelbaum:2011md}
  E.~Epelbaum, H.~Krebs, D.~Lee and U.-G.~Mei{\ss}ner,
  Phys.\ Rev.\ Lett.\  {\bf 106} (2011) 192501
  [arXiv:1101.2547 [nucl-th]].

\bibitem{Freer:2009zz} 
  M.~Freer {\it et al.}, 
  Phys.\ Rev.\ C {\bf 80}, 041303 (2009)

\bibitem{Zimmerman:2013cxa} 
  W.~R.~Zimmerman {\it et al.}, 
  arXiv:1303.4326 [nucl-ex].

\bibitem{Bijker:2000fw} 
  R.~Bijker and F.~Iachello,
  Phys.\ Rev.\ C {\bf 61}, 067305 (2000)

\bibitem {Kragh}
 H.~Kragh, Arch. Hist. Exact Sci. \textbf{64}, 721 (2010

\bibitem{Schellekens:2013bpa} 
  A.~N.~Schellekens,
  arXiv:1306.5083 [hep-ph].

\bibitem{Schlattl:2003dy}
  H.~Schlattl, A.~Heger, H.~Oberhummer, T.~Rauscher and A.~Csoto,
  Astrophys.\ Space Sci.\  {\bf 291} (2004) 27

\bibitem{Epelbaum:2012iu} 
  E.~Epelbaum, H.~Krebs, T.~A.~L\"ahde, D.~Lee and U.-G.~Mei{\ss}ner,
  Phys.\ Rev.\ Lett.\  {\bf 110}, 112502 (2013)

\bibitem{Epelbaum:2013wla} 
  E.~Epelbaum, H.~Krebs, T.~A.~L\"ahde, D.~Lee and U.-G.~Mei{\ss}ner,
  Eur. Phys. J. {\bf A 49}:82 (2013)

\bibitem{Epelbaum:2001fm} 
  E.~Epelbaum, U.-G.~Mei{\ss}ner, W.~Gloeckle and C.~Elster,
  Phys.\ Rev.\ C {\bf 65}, 044001 (2002)

\bibitem{Berengut:2013nh} 
  J.~C.~Berengut, E.~Epelbaum, V.~V.~Flambaum, C.~Hanhart, U.-G.~Mei{\ss}ner, 
  J.~Nebreda and J.~R.~Pelaez,
  Phys.\ Rev.\ D {\bf 87}, 085018 (2013)

\bibitem{Weinberg}
S.~Weinberg, {\em Facing up}, Harvard University Press (2001).

\bibitem{Pieper:1992gr} 
  S.~C.~Pieper, R.~B.~Wiringa and V.~R.~Pandharipande,
  Phys.\ Rev.\ C {\bf 46}, 1741 (1992)

\bibitem{Ishikawa:2007zz} 
  S.~Ishikawa and M.~R.~Robilotta,
  Phys.\ Rev.\ C {\bf 76}, 014006 (2007)

\bibitem{Bernard:2007sp} 
  V.~Bernard, E.~Epelbaum, H.~Krebs and U.-G.~Mei{\ss}ner,
  Phys.\ Rev.\ C {\bf 77}, 064004 (2008)

\bibitem{Bernard:2011zr} 
  V.~Bernard, E.~Epelbaum, H.~Krebs and U.-G.~Mei{\ss}ner,
  Phys.\ Rev.\ C {\bf 84}, 054001 (2011)

\bibitem{xx} 
R.~Bijker, J. Phys.: Conf. Ser. {\bf 380}, 012003 (2012)

\bibitem{Rupak:2013aue} 
  G.~Rupak and D.~Lee,
  arXiv:1302.4158 [nucl-th]

\bibitem{Navratil:2011zs} 
  P.~Navratil and S.~Quaglioni,
  Phys.\ Rev.\ Lett.\  {\bf 108}, 042503 (2012)

\bibitem{Luscher:1990ux} 
  M.~L\"uscher,
  Nucl.\ Phys.\ B {\bf 354}, 531 (1991).

\bibitem{Bour:2011ef} 
  S.~Bour, S.~Koenig, D.~Lee, H.-W.~Hammer and U.-G.~Mei{\ss}ner,
  Phys.\ Rev.\ D {\bf 84}, 091503 (2011)

\bibitem{Davoudi:2011md} 
  Z.~Davoudi and M.~J.~Savage,
  Phys.\ Rev.\ D {\bf 84}, 114502 (2011)

\bibitem{Bour:2012hn} 
  S.~Bour, H.-W.~Hammer, D.~Lee and U.-G.~Mei{\ss}ner,
  Phys.\ Rev.\ C {\bf 86}, 034003 (2012)

\bibitem{Rokash}
A.~Rokash {\it et al.}, {\em in preparation}.

\bibitem{Epelbaum:2008vj} 
  E.~Epelbaum, H.~Krebs, D.~Lee and U.-G.~Mei{\ss}ner,
  Eur.\ Phys.\ J.\ A {\bf 40}, 199 (2009)

%
\bibitem{Kaiser:2001jx} 
  N.~Kaiser, S.~Fritsch and W.~Weise,
  Nucl.\ Phys.\ A {\bf 697}, 255 (2002)

\bibitem{Lacour:2009ej} 
  A.~Lacour, J.~A.~Oller and U.-G.~Mei{\ss}ner,
  Annals Phys.\  {\bf 326}, 241 (2011)

\bibitem{Gezerlis:2013ipa} 
  A.~Gezerlis, I.~Tews, E.~Epelbaum, S.~Gandolfi, K.~Hebeler, A.~Nogga and A.~Schwenk,
  Phys.\  Rev.\  Lett.\  111, {\bf 032501} (2013)

\end{thebibliography}
\end{document}